\documentclass[twocolumn,groupedaddress,superscriptaddress,showpacs,prl,letterpaper]{revtex4}

\usepackage{times,amsmath,amssymb}      
\usepackage{graphicx} 
\usepackage{bm} 

\begin{document}

\title{Dirac Fermion Confinement in Graphene}

\pacs{81.05.Uw,71.10.-w,71.55.-i}

\date{\today}

\author{N.~M.~R. Peres}
\affiliation{Center of Physics and Departamento de F{\'\i}sica,
Universidade do Minho, P-4710-057, Braga, Portugal}
\author{A.~H. Castro Neto}
\affiliation{Department of Physics, Boston University, 590 
Commonwealth Avenue, Boston, MA 02215, USA}
\author{F. Guinea}
\affiliation{Instituto de  Ciencia de Materiales de Madrid, CSIC,
 Cantoblanco E28049 Madrid, Spain}

\begin{abstract}
We study the problem of Dirac fermion confinement in graphene in the
presence of a perpendicular magnetic field $B$. We show, analytically and numerically, that confinement leads to anomalies in the electronic spectrum and to a magnetic field dependent crossover from $\sqrt{B}$,
characteristic of Dirac-Landau level behavior, to linear in $B$ behavior,
characteristic of confinement. This crossover occurs when the radius of the
Landau level becomes of the order of the width of the system. As a result,
we show that the Shubnikov-de Haas oscillations also change as a function
of field, and lead to a singular Landau plot. We show that our theory
is in excellent agreement with the experimental data.
\end{abstract}

\maketitle

The production of two-dimensional (2D) graphene \cite{Netal04_short,Netal05b_short}, 
and the confirmation, via an anomalous integer quantum Hall effect
\cite{Netal05_short,Zetal05b_short}, of the presence of Dirac particles in
its electronic spectrum, has attracted a great deal of interest. Because of
the vanishing of the density of states at the Dirac point, these semi-metallic
systems present properties that deviate considerably from Landau's Fermi
liquid theory \cite{PGN05_short,Petal05_short}. In fact, these systems
show properties that are similar to models in particle physics
and, in particular, to relativistic quantum electrodynamics (QED) but with 
an effective "speed of light" (the Fermi-Dirac velocity, $v_F$) that is 
substantially smaller than the actual speed of light, $c$ ($v_F \approx
10^{-2} c$). In the most general case, the electron dispersion in
graphene can be written in the form of Einstein's equation: 
$E_{\pm}(k) = \pm \sqrt{m^2 v_F^4 + v_F^2 k^2}$, where $k$ is the electron
momentum (from now on we use units such that $c= \hbar = 1 = k_B$) and $m$ 
is the relativistic mass. In solids this mass represents a gap,
$\Delta = m v^2_F$, in the electronic spectrum. This gap can be generated, for instance, by 
the spin-orbit coupling \cite{kane}. 

Furthermore, due to experimental constraints, graphene samples are usually mesoscopic in size \cite{Betal04_short,Zetal05_short} leading to a situation where Dirac fermions
are confined by either zig-zag or armchair edges to a finite region in space  \cite{PGN05c_short}. Confinement is also particularly important
for the production of electron wave-guides that are the main elements
for the production of electronic devices such as all-carbon transistors. 
Dirac fermion confinement was a particularly enigmatic problem 
in the early days of quantum mechanics since the formation of wave packets in 
a region of the size of the Compton wavelength, $\approx 1/m$, requires the
use of negative energy solutions, or anti-particles, leading to a ground state 
with time dependent currents, the phenomenon called zitterbewegung
\cite{zuber}. Another manifestation of this confinement effect is Klein's paradox where a
flux of particles incident on a square potential barrier produces a
reflected current that is larger than the incident one. 

In this paper we show that graphene's zitterbewegung can be studied directly 
with the application of a transverse magnetic field. We show that the
confinement, generated by the finite size of the sample, shows up in a rather 
non-trivial way in the electronic spectrum. In particular, we show that the 
so-called Landau plots (the dependence of electronic spectrum in the magnetic 
field \cite{houten}) is rather non-trivial when the cyclotron length 
becomes of the order of the size of the sample. We address this problem
analytically by studying the Dirac equation in a magnetic field and also 
by solving numerically the tight-binding model for graphene 
in a finite geometry. 

Graphene is a honeycomb lattice of carbon atoms (with two sublattices, $A$
and $B$) with one electron per $\pi$-orbital (half-filled band) and can be
described by a tight-binding Hamiltonian of the form:
\begin{eqnarray}
H_{{\rm t.b.}} &=& - \sum_{\langle i,j \rangle ,\alpha} t_{ij} (a^{\dag}_{i,\alpha} b_{j,\alpha}
+ {\rm h.c.}) \, ,
\label{Htb}
\end{eqnarray} 
where $a^{\dag}_{i,\alpha}$ ($a_{i,\alpha}$) creates (annihilates) an
electron on site ${\bf R}_i$, with spin $\alpha$ 
($\alpha = \uparrow,\downarrow$), on sub-lattice $A$,  $b^{\dag}_{i,\alpha}$ ($b_{i,\alpha}$)  creates (annihilates) an
electron on site ${\bf R}_i$, with spin $\alpha$, 
on sub-lattice $B$, $t_{ij} = t
\exp\{i \theta_{ij}\}$  is the nearest
neighbor hopping energy ($t \approx 2.7$ eV) in the presence of a magnetic
field ${\bf B} = B \, {\bf z}$ ($\theta_{ij}=2\pi\int_i^j \bm A\cdot
d\bm{l}/\Phi_0$, 
with ${\bf A} = B \, x \, {\bf y}$ and $\Phi_0=2 \pi/e$ is the quantum of
magnetic flux). In the absence of next-nearest neighbor hopping, $t'$
($\approx 0.1 t$), the Hamiltonian is particle-hole symmetric \cite{PGN05_short} (the Zeeman energy is disregarded).

\begin{figure}[ht]
\begin{center}
\includegraphics*[width=7.cm,angle=-90]{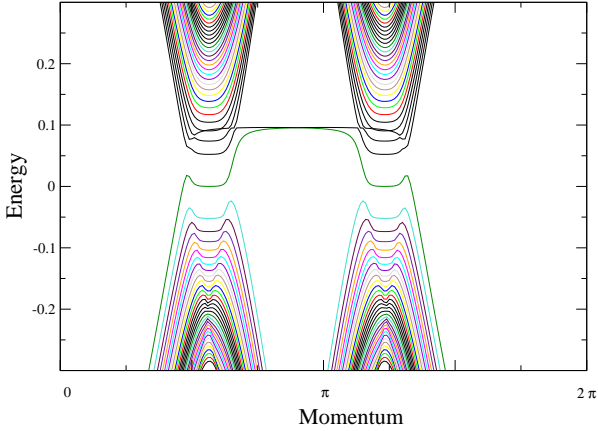}
\end{center}
\caption{(color on line) Energy spectrum (in units of $t$) for a graphene ribbon $600 a$ wide, as a function of the momentum $k$ along the ribbon (in units of $1/(\sqrt{3} a)$), with a magnetic flux of $5 \times 10^{-4} \Phi_0$ per hexagon, in the presence of confining 
potential (see text).}
\label{hifield}
\end{figure}

In a finite system,
one has to add the confining potential: 
$
H_e = \sum_i V_i n_i \, ,
$
where $n_i$ is the local electronic density. 
$V_i$ vanishes in the bulk but becomes large at the edge of
the sample. We have studied different types of potentials (hard wall,
exponential, and parabolic \cite{next}) but in this paper we will focus on a potential
that decays exponentially away from the edges into the bulk with a penetration depth, $\lambda$.  In Fig.~\ref{hifield} we show  the electronic spectrum for a graphene ribbon of width $L=600 a$ ($a =1.42$ \AA, is the Carbon-Carbon distance), in the presence of a
confining potential, $V(x) = V_0 \left[ e^{ -( x-L/2 ) / \lambda} + e^{ -( L/2-x ) / \lambda} \right]$, 
with strength $V_0 = 0.1 t$, and a penetration depth $\lambda = 150 a$ (we choose this large value of $\lambda$ just to illustrate the
effect of the confining potential in detail, in real samples we expect $\lambda \approx a$, which is the case discussed in the text), as a function of the momentum
$k$ along the ribbon. One can clearly see that in the presence of the confining potential the particle-hole symmetry is broken and, for $V_i>0$, the hole part of the spectrum
is strongly distorted. In particular, for $k$ close to the Dirac point, we see
that the hole dispersion is given by: 
$E_{n,\sigma=-1}(k) \approx -\gamma_n k^2 - \beta_n k^4$ where
$n$ is a positive integer, and $\gamma_n <0$ ($\gamma_n>0$) for $n<N^*$ ($n>N^*$). Hence, at $n=N^*$ the hole effective mass
{\it diverges} ($\gamma_{N^*} =0$) and, by tuning the chemical potential,
$\mu$, via a back gate, to the hole region of the spectrum ($\mu<0$) one
should be able to observe an anomaly in the Shubnikov-de Haas (SdH)
magneto-transport oscillations.

At low energies and long wavelengths, the energy spectrum of
Hamiltonian (\ref{Htb}) reduces to two Dirac cones centered
at the $K$ and $K'$ points in the Brillouin zone. Around 
each Dirac point the Hamiltonian (\ref{Htb}) can be written as:
\begin{eqnarray}
H_0 = \bm\sigma\cdot (v_F \bm p + e\bm A)  \, ,
\label{dirac}
\end{eqnarray}
where $v_F = 3 t a/2$,
$\bm\sigma$ are Pauli matrices acting on the states, $\psi({\bf r}) =
(\psi_A({\bf r}),\psi_B({\bf r}))$, 
of the two sub-lattices, and ${\bf p} = (p_x,p_y)= -i \bm\nabla$ is the 2D
momentum operator. 
In the absence of confinement ($V(x)=0$) we can diagonalize (\ref{dirac}) and one finds 
Landau levels given by: 
\begin{eqnarray}
E_{n,\sigma} &=& \sigma \sqrt{2} v_F \sqrt{n} \, \, /\ell_B 
= \sigma v_F \sqrt{2 e B \, n} \, ,
\label{ll}
\end{eqnarray}
where $\ell_B = 1/\sqrt{e B}$ is the cyclotron length,
$n$ is a positive integer, and $\sigma=1$ $(-1)$ labels the electron
(hole) levels.  

As discussed in the context of neutrino billiards \cite{berry_87}, 
the problem in the continuum suffers from the difficulty that
in trying to confine massless Dirac particles in a region of
size $L$ by including a large potential $V$ at the edge, leads to
a situation where particles still exist even at energies higher than
$V$. This problem, of course, does not arise in the tight-binding
description. In order to avoid this problem in the continuum
description, we introduce a position dependent mass term:
$
H_c = v_F^2 M(x) \sigma_z \, ,
$
where, 
\begin{equation}
M(x) =
\left\{
\begin{array}{c}
M , \hspace{1cm} x<-L/2 \,, \\
0 , \hspace{0.4cm} -L/2 \leq x \leq L/2 \,, \\
M , \hspace{1cm} x>L/2 \,, \\
\end{array}
\right.
\end{equation}
where $L$ is the width of the graphene stripe. We are interested 
in the hard wall case ($M \to \infty$) although other potentials
can be studied in analogous way \cite{next}. Notice that in the
absence of an applied magnetic field, a mass term does not break 
particle-hole symmetry, as in the case of a potential
$V(x)$)\cite{berry_87}.  Nevertheless, since both $V(x)$ and $M(x)$ 
are strongly concentrated at the edge (in a distance of the order
of the lattice spacing), they do not modify the states
in the bulk. It is also worth mentioning that although (\ref{dirac}) is not
time reversal symmetric in the absence of a magnetic field \cite{berry_87}, 
time reversal is recovered by the inclusion of the second Dirac cone at the 
opposite side of the Brillouin zone.

\begin{figure}[ht]
\begin{center}
\includegraphics*[width=8cm]{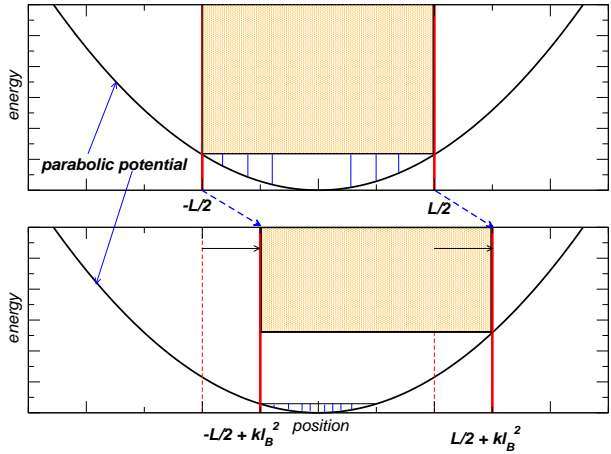}
\end{center}
\caption{(color on line) Illustration of the potentials in the problem:
$k=0$ (top) and $k\ne 0$ (bottom).}
\label{fwkb}
\end{figure}

The Dirac equation, $H \psi = E \psi$, 
where $H=H_0+H_c$, can be recast in terms of a wavefunction ansatz, $\phi = (H+E)\psi$,
as: $H^2 \phi = E^2 \phi$. It is easy to show that this wavefunction
has the form:
$
\phi(x,y) \propto e^{-i k y} \varphi_{\sigma}(x) \, ,
$
where $k$ is the momentum along the $y$ direction,
$\sigma=\pm 1$ are the eigenstates of $\sigma_z$, and
$\varphi_{\sigma}(x)$ obeys the following equation:
\begin{eqnarray}
\left[-\frac{\partial^2}{\partial \xi^2} + \xi^2 + V(\xi)\right] \varphi_{\sigma}(\xi) = \epsilon \varphi_{\sigma}(\xi) \, 
\label{main}
\end{eqnarray} 
where,
\begin{eqnarray}
\xi &=& x/\ell_B - k \ell_B \, ,
\label{xi}
\\
V(\xi)&=& [v_F \ell_B M(\xi)]^2 \, ,
\label{vxi}
\\
\epsilon &=& (\ell_B E_{\sigma}/v_F)^2-\sigma \, .
\label{epsig}
\end{eqnarray}
Equation (\ref{main}) is a dimensionless Schr\"odinger equation
for a non-relativistic particle (of mass $1/2$) in a
parabolic potential (of frequency $\omega_0=2$) superimposed
to a potential well, $V(\xi)$, whose position shifts with the momentum $k$ 
(see Fig.~\ref{fwkb}). We see, from (\ref{epsig}), that the Dirac fermion spectrum 
in the presence of the magnetic field and confining potential can be written as:
\begin{eqnarray}
E_{\epsilon,\sigma} = \sigma v_F \sqrt{\epsilon+\sigma}/\ell_B \, .
\label{spec}
\end{eqnarray}

At low energies (small $\epsilon$) 
the parabolic potential dominates and the wavefunctions look like the ones 
in the infinite system ($L \to \infty$) in the presence of magnetic field \cite{PGN05_short}. 
In this limit, the spectrum of (\ref{main}) is given by the 1D harmonic
oscillator: $\epsilon \approx \omega_0(n+1/2) = 2 n + 1$. This result,
together with (\ref{spec}), gives 
rise to eq.~(\ref{ll}), with the energy proportional to $\sqrt{B}$. 
On the other hand, at large energies, the confining 
potential becomes more important and the energy spectrum changes to: $\epsilon \approx 
[\pi n/(L/\ell_B)]^2$, leading to a Dirac fermion spectrum of the
form:
\begin{eqnarray}
E_{n,\sigma} \approx 
\sigma v_F [\pi n/L + \sigma L e B/(2 \pi n)] \, ,
\label{conf}
\end{eqnarray}
which leads to a spectrum proportional to $B$. 
These simple arguments show that the SdH
magneto-resistance oscillations changes behavior as a function of magnetic
field. On the one hand, for a given chemical potential $\mu$, eq.~(\ref{ll})
predicts that the maxima of the SdH should happen at fields:
\begin{eqnarray}
1/B_F(N,L) = (2 v_F^2 e/\mu^2) \, N \, ,
\label{bfn1}
\end{eqnarray}
where $N$ is the Landau level index. On the other hand, eq.~({\ref{conf}) 
shows that the maxima occur at fields:
\begin{eqnarray}
1/B_F(N,L) = [L^2 e/(2 \pi^2 N)]/(N_c(L)-N) \, ,  
\label{bfn2}
\end{eqnarray}
where 
\begin{eqnarray}
N_c(L) = \mu L/(\pi v_F) \, .
\label{ncl}
\end{eqnarray}
Hence, $1/B_F(N)$ {\it diverges} at a critical Landau 
level index $N_c(L)$ which increases linearly with the width $L$ of
the graphene stripe. The deviation from (\ref{bfn1}) to (\ref{bfn2})
is a clear sign of the Dirac fermion confinement.

\begin{figure}[hpt]
\begin{center}
\includegraphics*[width=8cm]{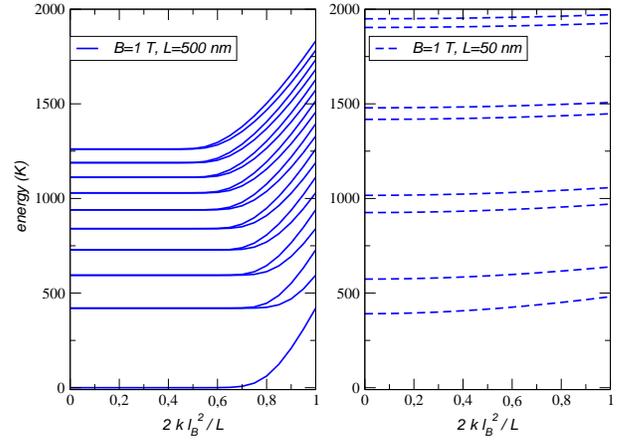}
\end{center}
\caption{(color on line) Energy spectrum, $E_\sigma(k)$,
as function of $k \ell_B^2/L$ for two system sizes, $L=50$ nm and
$L=500$ nm, at $B=1$ T, for 10 eigenstates. 
Each Landau level is sub-divided into two sub-levels, distinguished
be the quantum number $\sigma$.}
\label{fig_energy_levels}
\end{figure}

Notice that the crossover from (\ref{ll}) to (\ref{conf}) 
(or from (\ref{bfn1}) and (\ref{bfn2})) occurs when the Landau orbit fits into
the confining potential. Since each orbit must enclose exactly an
integer number of flux quantum, $\Phi_0$,  
the crossover occurs at a magnetic field $B^*$ such that, $B^* L^2 \approx N \Phi_0$, that is,
\begin{eqnarray}
B^*(N,L) = N \Phi_0/L^2 \, .
\label{bstar}
\end{eqnarray} 

Let us now consider the numerical solution of the 
differential equation (\ref{main}) written in terms
of the above introduced dimensionless variables in the case
$M\rightarrow\infty$. Because our treatment of the Dirac equation
leads to a second order differential equation, the appropriate boundary 
conditions for a sharp confining edge with a mass term is \cite{MF04}: $\varphi_{\sigma} (-L/(2 \ell_B)+ k \ell_B)=\varphi_{\sigma}(L/(2 \ell_B)+ k \ell_B)=0$.
In Fig.~\ref{fig_energy_levels} we show the energy spectrum
at $B=1$ T for two different system sizes as a function of $k \ell^2/L$.
One can clearly see that the degeneracy of the Landau levels is lifted
for small enough system sizes or large enough $k$ \cite{BF06a,Abanin06_short}. 
For small $k$ the energy states are dispersionless and degenerate.

\begin{figure}[hpt]
\begin{center}
\includegraphics*[width=8cm]{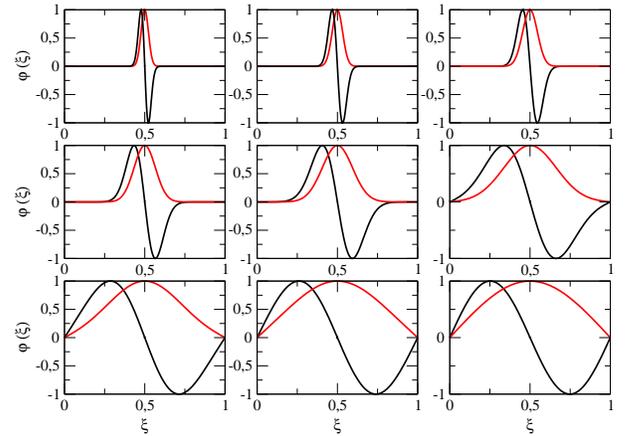}
\end{center}
\caption{(color on line) Wave functions $\varphi_{\sigma}$ for $L=500$ nm. 
The values of the magnetic field are, from left to right and top to bottom,
$B=$ 5, 2.5, 1.25, 0.6, 0.3, 0.1, 0.05, 0.025, 0.01 (T). } 
\label{fig_wave_function}
\end{figure}

In Fig.~\ref{fig_wave_function} we show the first two state eigenvalues of 
the effective Schr\"odinger equation (\ref{main}), for $k=0$ and
different values of the field (or, equivalently, different system
sizes). One can clearly observe the change in the wavefunction
from cosine (sine) to gaussian (first order Hermite polynomial
times gaussian) behavior as $l_B$ decreases. Clearly the system
evolves from a state where the boundaries introduced by the 
confining potential
are irrelevant (the wave functions and the energy levels
are essentially those of the 1D harmonic oscillator at $B=5$ T,
and $l_B\simeq$ 12 nm), passing to a state where the gaussian decay of the wave 
function in the classically forbidden regions is important 
allowing the electrons to feel the presence of the confinement potential
(the wave functions and the energy levels cannot be described either by the 1D harmonic  oscillator or by the particle in a box for $B=0.1$ T at $l_B\simeq$ 80 nm). Finally, when the
Landau orbit is of the order of size of the confinement potential, 
the eigenstates are essentially those of the particle in a box 
($B=0.01$ T, $l_B\simeq$ 250  nm).  

\begin{figure}[hpt]
\begin{center}
\includegraphics*[width=8cm]{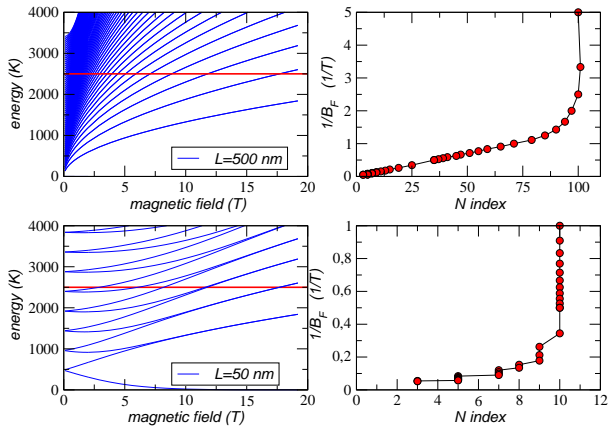}
\end{center}
\caption{\label{fig_van_Houten}(color on line)  
Landau plots for two system sizes,
$L$= 500, 50 nm. On the right hand side, we depict the energy
levels (in K) as a function of magnetic field (in T) and the horizontal
line marks the position of the Fermi energy. On the left hand size
we plot $1/B_F(T)$ as a function of the Landau index $N$. } 
\end{figure}

In Fig.~\ref{fig_van_Houten}, we show the energy spectrum
as a function of the magnetic field for different system sizes
together with their respective Landau plots. Note that at small
fields (when $L$ is large enough) the energy spectrum follows the $\sqrt{B}$ dependence
of (\ref{ll}) while at larger fields it becomes linear in $B$ as
predicted by (\ref{conf}). The crossover from these two asymptotic
behaviors is indeed given by eq.~(\ref{bstar}), as one can see
from the size dependence. More striking, however, is that fact
that $1/B_F$ indeed diverges at sufficiently
high Landau level index and that the size dependence is given
by (\ref{ncl}). 

\begin{figure}[hpt]
\begin{center}
\includegraphics*[width=8cm]{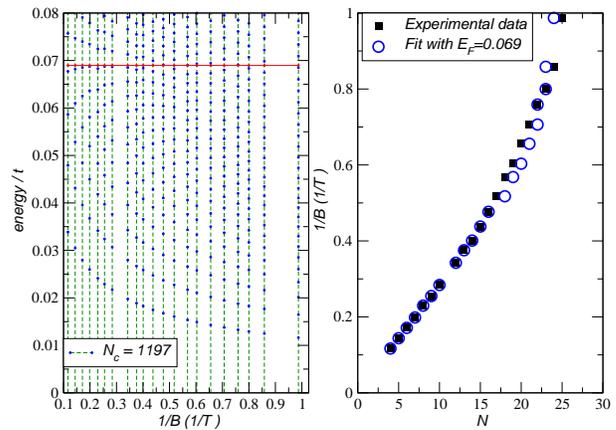}
\end{center}
\caption{\label{comparison}(color on line)  On the left hand side,
we show the energy spectrum (in units of $t$) as a function
of $B^{-1}$ (in T$^{-1}$). The horizontal line shows position of
the Fermi energy ($\approx 0.069 t$). On the right hand side, 
we show the theoretical Landau plot (open circles) in comparison
with the experiments of ref.~[\onlinecite{Betal06_short}] (close squares).}
\end{figure}

In Fig.~\ref{comparison} we compare our tight binding results with with the experimental data of ref.~[\onlinecite{Betal06_short}]. We choose
a ribbon of size $295$ nm (equivalent to $N_c=1197$ unit cells),
and Fermi energy $0.069 t$ (equivalent to $0.22$ eV). Notice
the excellent agreement between theory and experiment for $N<16$
($2{\rm T}<B<10{\rm T}$). For $N>16$ there is shift of $N$ by one (either plus or minus one) relative to the experiment. This discrepancy, we believe, can be assigned to the experimental difficulty in assigning the Landau indices at small magnetic fields \cite{Betal06_short}.

In summary, we have studied the problem of Dirac confinement in graphene,
that is, graphene's zitterbewegung, 
for graphene stripes of size $L$ in the presence of a transverse magnetic
field, $B$. We show that the interplay between size effects and magnetic
field can be studied in the continuum limit using the Dirac equation coupled 
to a vector potential. We present arguments that show that the 
spectrum of the problem shows a crossover from magnetic field dominated to 
confinement dominated as a function of magnetic field or system size. 
The crossover occurs when the radius of the Landau level becomes of the order of the width of the system. In the crossover the 
spectrum changes from $\sqrt{B}$ to linear in $B$ and that the Landau plots, 
that can be measured in a SdH experiment, change from dramatically in the 
presence of a finite system. Our results are in excellent agreement with experiments.

We are very grateful to C. Berger and W. A. de Heer 
for providing the experimental data. 
A.H.C.N. was supported through NSF grant DMR-0343790. N. M. R. P. thanks 
ESF Science Programme INSTANS 2005-2010, and FCT under the grant POCTI/FIS/58133/2004. F. G. acknowledges
funding from MEC (Spain) through grant FIS2005-05478-C02-01.

\bibliography{graphite0_1}

\end{document}